\documentclass[12pt]{elsarticle}
\usepackage{booktabs}
\usepackage{graphicx}
\usepackage{float}   %fixe les images avec [H] dans figure
\usepackage{array}
\usepackage{amssymb}
\usepackage[pdftex]{thumbpdf}
\usepackage{epstopdf}
\usepackage{multirow}
\usepackage{textcomp}
\usepackage{dcolumn}
\newcolumntype{d}[1]{D{.}{.}{#1} }
\usepackage[utf8]{inputenc}
\makeatletter
\def\ps@pprintTitle{%
 \let\@oddhead\@empty
 \let\@evenhead\@empty
 \def\@oddfoot{}%
 \let\@evenfoot\@oddfoot}
\makeatother
\begin{document}

\begin{frontmatter}

\title{Final Results of the PICASSO Dark 
Matter Search Experiment}

\author[iusb]{E.~Behnke} 
\author[qu]{M.~Besnier} 
\author[saha]{P.~Bhattacharjee} 
\author[qu]{X.~Dai\fnref{chalk}}
\author[saha]{M.~Das} 
\author[qu]{A.~Davour} 
\author[mtl]{F.~Debris}
\author[lu]{N.~Dhungana}
\author[lu]{J.~Farine} 
\author[mtl]{M.~Fines-Neuschild}
\author[ab]{S.~Gagnebin} 
\author[qu]{G.~Giroux} 
\author[iusb]{E.~Grace\fnref{sterl}} 
\author[mtl]{C.~M.~Jackson\fnref{berkley}} 
\author[qu]{A.~Kamaha}
\author[ab]{C. B.~Krauss} 
\author[mtl]{M.~Lafreni\`ere}
\author[mtl]{M.~Laurin}
\author[snolab]{I.~Lawson} 
\author[mtl]{L.~Lessard} 
\author[iusb]{I.~Levine} 
\author[ab]{D.~Marlisov} 
\author[mtl]{J.~-P.~Martin} 
\author[ab]{P.~Mitra} 
\author[qu]{A.~J.~Noble} 
\author[mtl]{A. Plante\corref{cor}}
\author[lu]{R.~Podviyanuk} 
\author[prague]{S.~Pospisil} 
\author[lu]{O.~Scallon} 
\author[saha]{S.~Seth\fnref{Tata}}
\author[mtl]{N.~Starinski} 
\author[prague]{I.~Stekl} 
\author[lu]{U.~Wichoski} 
\author[mtl]{V.~Zacek}

\address[iusb]{Department of Physics \& Astronomy, Indiana University South Bend, South Bend, IN 46634, USA}
\address[qu]{Department of Physics, Queen's University, Kingston, K7L 3N6, Canada}
\address[saha]{Saha Institute of Nuclear Physics, Centre for AstroParticle Physics (CAPP), Kolkata, 700064, India}
\address[mtl]{D\'epartement de Physique, Universit\'e de Montr\'eal, Montr\'eal, H3C 3J7, Canada}

\address[lu]{Department of Physics, Laurentian University, Sudbury, P3E 2C6, Canada}
\address[ab]{Department of Physics, University of Alberta, Edmonton, T6G 2G7, Canada}
\address[snolab]{SNOLAB, Lively ON, P3Y1N2, Canada}
\address[prague]{Institute of Experimental and Applied Physics, Czech Technical University in Prague, Prague, Cz-12800, Czech Republic}

\fntext[chalk]{Present address: AECL Chalk River Laboratories, Chalk River, K0J 1J0, Canada}
\fntext[berkley]{Present address: Univ. of California, Berkeley, CA 94720, USA}

\fntext[Tata]{Present address: Tata Institute of Fundamental Research, Mumbai, 400088 Maharashtra, India}

\fntext[sterl]{Present address: Sterling College, Dep. of Natural Sciences \& Mathematics, Sterling, KS 67579, USA }

\cortext[cor]{Corresponding author: e-mail: arthur.plante@umontreal.ca}

\begin{abstract}

The PICASSO dark matter search experiment operated an array of 32 superheated droplet detectors containing 3.0 kg of C$_{4}$F$_{10}$ and collected an exposure of 231.4 kgd at SNOLAB between March 2012 and January 2014. We report on the final results of this experiment which includes for the first time the complete data set and improved analysis techniques including \mbox{acoustic} localization to allow fiducialization and removal of higher activity regions within the detectors. No signal consistent with dark matter was observed. We set limits for spin-dependent interactions on protons of $\sigma_p^{SD}$~=~1.32~$\times$~10$^{-2}$~pb (90\%~C.L.) at a WIMP mass of 20 GeV/c$^{2}$. In the spin-independent sector we exclude cross sections larger than $\sigma_p^{SI}$~=~4.86~$\times$~10$^{-5 }$~pb~(90\% C.L.) in the region around 7 GeV/c$^{2}$. The pioneering efforts of the PICASSO experiment have paved the way forward for a next generation detector incorporating much of this technology and experience into larger mass bubble chambers.
\end{abstract}

\begin{keyword}
dark matter \sep WIMPs \sep superheated droplets \sep SNOLAB

\end{keyword}

\end{frontmatter}
\section{Introduction}
\label{Int}

 Dark matter searches are the focus of underground laboratories all over the world. Even though the existence of dark matter is no longer \mbox{controversial}, the particle nature of dark matter has not been established so far. The class of particles best motivated theoretically are usually referred to as WIMPs, or Weakly Interacting Massive Particles \cite{PDG,Supersymmetric_dark_matter,DM_Detection}. The experimental signature of such particles can be searched for in production experiments at colliders and beam dumps, indirectly by looking for annihilation products from zones expected to have high dark matter densities, such as the galactic core, or by directly looking for interactions between ordinary matter and dark matter through the observation of nuclear recoils in large underground detectors. 
\\
\indent The direct detection of dark matter through the observation of nuclear recoils requires detector technologies sensitive to keV nuclear recoils while able to discriminate against abundant backgrounds from conventional radioactivity. Successful technologies have been developed based on cryogenic solid state detectors, scintillating crystals, noble liquids and superheated liquids \cite{SNOWMASS}. Historically, the interaction of dark matter with normal matter has been divided into two categories, spin independent and spin dependent.  Since theory provides little guidance on WIMP masses or their couplings it is important to explore both sectors with a wide variety of targets.
\\
\indent The highest sensitivity in the spin independent sector has been obtained by experiments using noble liquids and cryogenic crystals (e.g. LUX, XENON, PandaX, CDMS \cite{LUX2016SI_Run3Run4,xenon2016,PandaX,SuperCDMS}). In the spin dependent (proton) sector the superheated detector technology has been at the forefront since several years, with the most stringent limits set by PICO (formed from a merger of \mbox{PICASSO} and COUPP) \cite{PICO2L_RUN2,PICO60_CF3I,CF3I_Calib_COUPP}. Other experiments using this technique are \mbox{SIMPLE} and MOSCAB \cite{SIMPLE,MOSCAB}. Two primary types of detectors are in use: droplet detectors and bubble chambers, all using fluorinated halocarbons as target liquids.
\\
\indent The PICASSO experiment at SNOLAB used a superheated liquid droplet target of C$_{4}$F$_{10}$. A fluorine rich target such as C$_{4}$F$_{10}$ is ideal for dark matter searches in the spin-dependent sector due to the very high spin enhancement factor from the single unpaired proton in $^{19}$F and its natural isotopic abundance of 100\% \cite{ref4,ref5}. The low mass number also leads to a peak sensitivity in the low WIMP mass range of tens of GeV/c$^{2}$, an area of much recent interest in dark matter experiments \cite{DAMA_LIBRA,CoGeNT,CDMS_SI_Signal}. In this mass region a competitive spin-independent search can also be performed.
\
%%%%%%%%%%%%%%%%%%%%%
\section{Detection Principle}
\label{Detection_Principle}
\begin{figure}[H]
\centering
\includegraphics[width=0.8\textwidth, origin=c, angle=0]{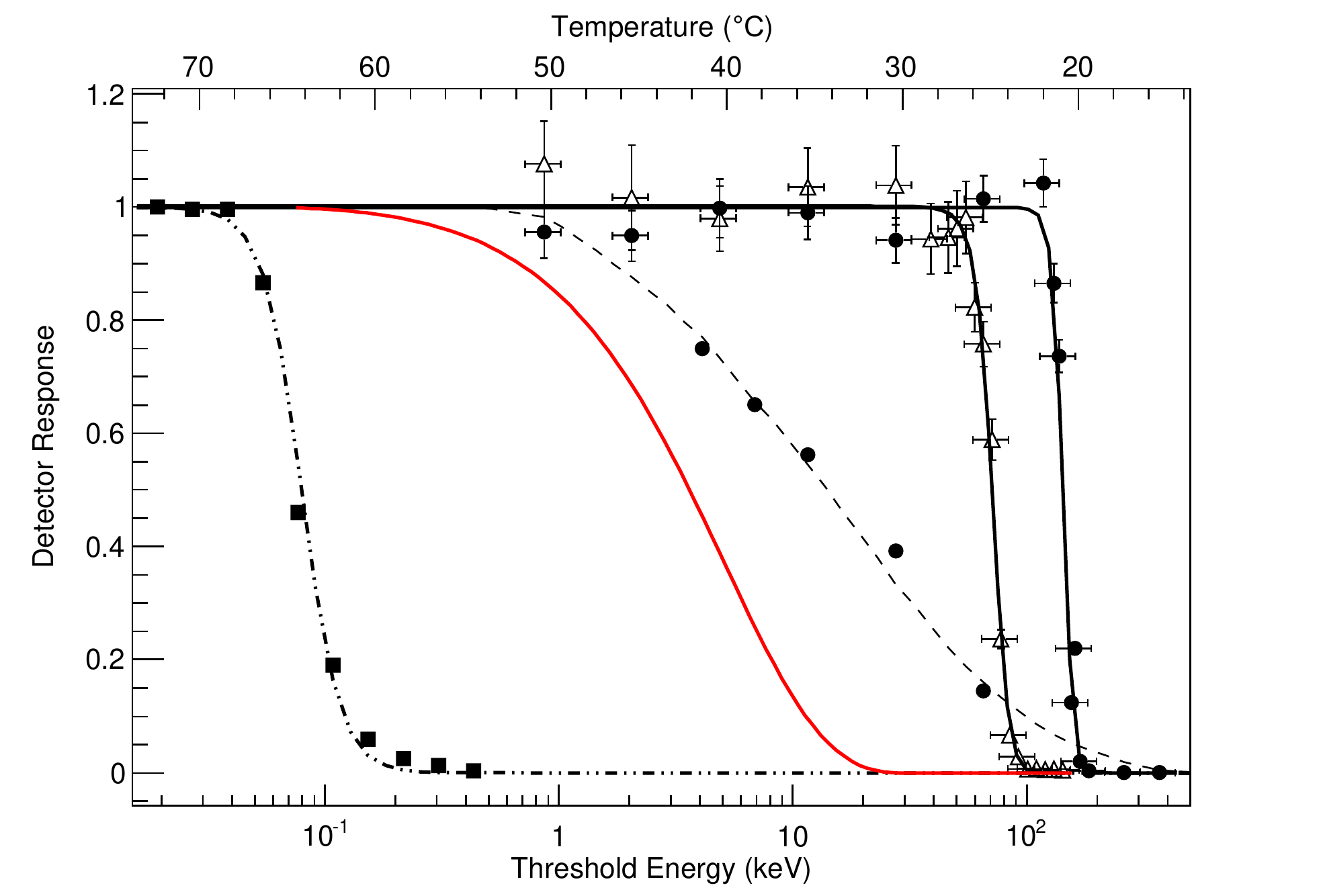}
\caption{Response to different kinds of particles in superheated C$_{4}$F$_{10}$. From left to right: 1.75 MeV $\gamma$-rays and MIP’s (dot-dashed); $^{19}$F recoils modeled assuming the scattering of a 50 GeV/c$^2$ WIMP (continuous red); poly-energetic neutrons from an AcBe source (dotted); $\alpha$ - particles at Bragg peak from $^{241}$Am decays (open triangles);  $^{210}$Pb recoil nuclei from $^{226}$Ra spikes (full dots).  For $^{19}$F and $^{210}$Pb-recoils this energy scale corresponds directly to their detection thresholds.}

\label{fig1}
\end{figure}

 The detection principle of PICASSO is a variant of the classical bubble chamber technique where a superheated liquid is held in a metastable state such that a deposition of a critical energy within a critical radius causes a phase transition and a droplet to change from liquid to gas \cite{ref9,ref10,ref11,ref12}. The explosive bubble nucleation is accompanied by an acoustic signal in the audible and ultrasonic frequency range and gives information on the nature of the underlying event \cite{CF3I_Calib_COUPP,ref13,PICASSO_CALIB_2011}. Since the detector observes phase transitions it performs as a threshold device, which can be controlled by setting the temperature and/or pressure.
 \\
\indent With a boiling temperature of \textit{T}$_{b}$ = $-$1.7~$^{0}$C at a pressure of 1.013 bar, the C$_{4}$F$_{10}$ droplets in PICASSO are kept in a moderately superheated state at temperatures from 25~-~50~$^{0}$C corresponding to thresholds in the range \mbox{1~-~60~keV}. The precise relation between energy threshold and operating temperature in C$_{4}$F$_{10}$ was determined by extensive measurements of \mbox{$^{19}$F-recoils} using mono-energetic neutron beams and with alpha emitters of known energies in the droplets  \cite{PICASSO_CALIB_2011,PICASSO2012,PICASSO_NIM}. Since each temperature corresponds to a defined \mbox{energy} threshold, the spectrum of the particle induced energy depositions can be reconstructed by varying the threshold temperature. A summary of the detector response to different kinds of particles is shown in Fig.\ref{fig1}, where temperatures are converted into energy thresholds. For  $^{19}$F-recoils this energy scale corresponds directly to their detection \mbox{thresholds}. 

\indent Since WIMP induced recoil energies of $^{19}$F nuclei are expected to be smaller than 100 keV they become detectable above 30~$^{0}$C.  At the normal operating thresholds of PICASSO above 1 keV, particles with low ionization densities, such as $\gamma$~-~rays and $\beta$~-~particles do not deposit sufficient energy to induce a phase change and these events are suppressed by more than a factor of  10$^{-9}$. Only alpha particles and neutrons can contribute particle induced backgrounds to the WIMP searches in this detector. The described responses depend exclusively on the thermodynamic parameters describing the degree of superheat of the droplet fluid and are independent of detector specific parameters (i.e. droplet size, loading fraction, transducer response).     
\\
\indent Since $\alpha$-particles induce phase transitions over the entire range of the WIMP sensitivity due to their large \textit{dE}/\textit{dx}, they are, together with neutrons an important background for this kind of detector in dark  matter searches. However the shapes of the WIMP response, with count rates decreasing with increasing threshold energy, and of the $\alpha$-response with constant rates in the region of \mbox{interest}, differ substantially, such that both contributions can be separated by fitting. In addition, PICASSO discovered that for alpha particles in the bulk fluid, it was possible to discriminate between alpha particles and nuclear recoils, an advantage best exploited in the next generation bubble chambers. A detailed discussion of the detector response is given in~\cite{PICASSO_CALIB_2011}.

%%%%%%%%%%%%%%%%%%%%%
\section{Detector Set-Up and Operation}
\label{Detector_Set-Up_and_Operation}

 PICASSO started operating 4.5 L volume droplet detectors at \mbox{SNOLAB} in 2007 and published data with increasing exposure and sensitivity in 2009 and 2012 \cite{PICASSO2009,PICASSO2012}. Being limited essentially  by the alpha background in the gel components, continuous efforts were made to revise and improve the purification procedures and to replace higher rate modules with cleaner ones (Section \ref{Acoustic_Triangulation}). Data taking with the upgraded set of 32 detector modules started in March~2012 and concluded in January~2014 with a total exposure of 231.4~kgd, after applying fiducial volume and timing cuts (Section \ref{Analysis}). This last run period was enhanced by the overall lower background rates and by an optimization of the data taking, concentrating on low energy threshold measurements between 1 and 2~keV.
\\
\indent The PICASSO detectors and their operation principle have been described in detail in \cite{PICASSO_CALIB_2011,PICASSO2012}. The current generation of detectors consisted of cylindrical modules of 17 cm diameter and 40 cm height. They were fabricated from acrylic and were closed on top by stainless steel lids sealed with polyurethane O-rings. The acrylic walls of the cylinders had a thickness of 1.3 cm in order to provide sufficient  mechanical strength and to minimise radon leakage. Each detector was filled with a water saturated polyacrylamide emulsion up to 30 cm in height and loaded with droplets of C$_{4}$F$_{10}$ with an average diameter of 200 \textmu m. The active part of each detector was topped by mineral oil, which is connected to a hydraulic manifold.
\\
\indent The emulsion of the droplets was created by a magnetic stirrer where the time and speed were adjusted to obtain a bell shaped droplet volume distribution centered on diameters of 200 \textmu m and with a distribution width of about 150 \textmu m (FWHM). The selected droplet size was found to maximize the amount of active fluid in the detectors. Much larger droplets would tend to imperil the structural integrity of the surrounding polymer during bubble formation in the gel; smaller droplets would increase the geometric efficiency for detection of alpha particles originating from the gel. Calibrations showed that the observed particle response was independent of the droplet size within the range considered and conformed to the response observed in bulk superheated liquids. Above 45~$^{0}$C the polymer becomes increasingly softer; non-particle induced phase transitions appear due to shear and fractures and these events become a non-negligible, but still controllable background (Section \ref{Acoustic_Triangulation}).
\\
\indent The initial active mass of each detector was known with a precision of 1\% from weighing during fabrication, but additional uncertainties arise due to potential losses during polymerization, diffusion into the gel matrix and surface leakage. Therefore the active detector mass and sensitivity were verified and monitored by measurements with a calibrated AmBe source at periodic intervals. No loss was observed over the run period defined, and the total mass of C$_{4}$F$_{10}$ in the set-up was determined to be 2.97~$\pm$~0.15 kg, corresponding to 2.37~$\pm$~0.12 kg of $^{19}$F.
\\
\indent The acoustic signal associated with an event was observed by 9 piezoelectric transducers (Ferroperm P27) uniformly distributed around each detector at three different heights on the container wall (one of the detectors had 6~piezos). This arrangement allowed the events to be localized by triangulation and the definition of fiducial volumes to avoid higher background regions near the container walls (Section \ref{Acoustic_Triangulation}). Triggering of any of the nine transducers  causes all channels to acquire data. The trigger is fully sensitive over the entire threshold range \cite{DAQ_PICASSO}. 
\\
\indent The detectors were typically operated for 40~-~50 hours before being compressed by a hydraulic system to prevent damage to the gel matrix due to slow and  continuing bubble growth. A compression phase which reduced the bubbles to the original liquid droplet state lasted 12 hours at a pressure of 6.2 bar. This relatively long compression time was selected to assure complete curing of the gel and had no effect on the droplet size distribution and sensitivity which would have shown up during calibration runs. 

The complete system of 32 detectors was housed in eight groups of four in thermally isolated boxes and was temperature controlled to roughly 0.1~$^{0 }$C to have a well-defined detection threshold. In order to preserve the sensitivity for annual rate modulations of an eventual signal, all detectors were operated at the same temperature at the same time. All 32 detectors met the data quality requirements and were used in the analysis. 
\\
\indent The entire installation was surrounded by 50 cm of water contained in polyethylene tanks which served as neutron moderator and shielding. This shielding and setup represented a significant reduction in the background activity due to neutrons compared to the underground shielding used for previous PICASSO data sets. At the SNOLAB facility almost all neutrons are produced via ($\alpha$, n) reactions due to natural U/Th radioactivity in the rock, with a remaining 10\% from fission. A production rate of 4.0~\mbox{neutrons}~g$^{-1}$y$^{-1}$ was found from the relative abundance of isotopes in the surrounding Norite rock by computations using the SOURCES code \cite{Alvine_thesis,alpha_n_reaction}. These neutrons were further propagated by a GEANT4 simulation through the rock, the cavern and the water shielding to the detector location \cite{geantref1,geantref2}. The performance of the simulation and the effectiveness of the shielding were checked by measurements with several $^{3}$He counters (SNO NCDs \cite{SNONCDs}) which were surrounded by various thicknesses of dedicated polyethylene neutron moderator \cite{Alvine_thesis}. Measurements and simulations with and without water shielding showed that 99.66~$\pm$~0.01~\% of the incoming neutrons with energies above 5~keV were stopped in the shielding. Using the estimate of the fast neutron flux underground of (4~$\pm$~2)$\times$ 10$^{3}$~neutrons~m$^{-2}$d$^{-1}$ in the cavern \cite{Bkg_SNOLAB} and an average sensitivity of PICASSO detectors to neutrons of 0.1~cts~per~\mbox{neutron}~g$^{-1}$cm$^{-2 }$~\cite{PICASSO_NIM}, the expected event rate induced by fast neutrons was determined to 0.14~cts~kg$^{-1}$d$^{-1}$. This rate is still more than a factor ten smaller than the sensitivities of the best detectors in the set up.
\\
\indent The signals of each piezoelectric sensor were digitized using custom electronics with a sampling rate of 800 kHz and 16384 samples per event. The data acquisition underwent a doubling of the sampling frequency since the previous runs \cite{PICASSO2012} with the goal of improving the ability to reject alpha background. This important feature discovered by PICASSO was however not sufficiently efficient to be useful in this analysis due to the only partial containment of alpha events in the droplets  \cite{ref13}. The definition of a good event was determined by cut parameters on five acoustic variables described below. To determine these cuts the array of detectors was calibrated with a weak poly-energetic AmBe neutron source (68.71~$\pm$~0.74~n~s$^{-1}$) at every temperature that had a significant exposure. These calibration data were spread over the entire data taking period in order to follow temporal variations of the event selection parameters. A total of 53.8 kgd worth of neutron data was acquired for the calibrations. 
 
\section{Acoustic Signatures for Background Discrimination}
\label{Acoustic_Signatures_for_Background Discrimination}

Calibrations with neutron test beams and fast neutrons from AcBe and AmBe sources showed that the waveforms associated with particle induced acoustic signals have characteristic frequency spectra (FFT) and time dependencies \cite{PICASSO_NIM}. The signals have a short rise time, reaching a maximum after 20~-~40~\textmu s, with slower oscillations following for several milliseconds. In addition, the amplitude distributions of the high frequency content ($>$~20~kHz) of the particle induced wave forms were concentrated in a well-defined peak. Tests on different known droplet samples showed that the amplitudes of particle induced events were not droplet size dependent, which is consistent with the current model of particle~-~induced sound generation in superheated liquids described in \cite{PICASSO_CALIB_2011}.
These features and others were used to construct variables which allowed the discrimination of particle induced events from non-particle background events. Since all event selection variables were dependent on the detector module and the operating temperature, a set of  variables and cut values were calculated using neutron calibration data averaged over all piezoelectric transducers. The cut values were determined by plotting the variable's distribution obtained in neutron calibrations and setting the cuts to retain 95\% of neutrons. The values obtained were then fit with a polynomial to interpolate to all operating temperatures.The following are the main variables for event discrimination: 

\indent \textit{ The acoustic energy EVAR} is a variable calculated using the integrated energy in the recorded waveforms. The variable resolution was improved from the previous publication by reducing the signal time window used to calculate the variable to a length of 500 \textmu s (starting 125 \textmu s before the event trigger). This variable primarily isolates particle induced events and removes electronic noise which tends to have less acoustic energy than bubble events~\cite{Alvine_thesis}.
\\
\indent\textit{ The signal rise time variable RVAR} is calculated by taking the standard deviation of the time bins in the first 100 \textmu s following the signal start time \cite{Alvine_thesis}. This primarily removes electronic noise and so-called ``mystery" events, described later, which have a characteristically slow rise time. 
\\
\indent \textit{ The event shape/quality variable QVAR} is calculated by taking the ratio of signal power within the first and second 10~ms of the recorded signal time window and this removes events where the signal power is distributed equally between the two time windows. These are events with unusual shapes, such as long ringing type signals, due to electronic noise.
\\
\indent The\textit{ event time variable TVAR } is calculated by finding the mean time bin of the signal squared and is used to identify events where the acoustic power is concentrated later in the signal. This was used to remove a class of repeating events due to delayed signals and electronic glitches. 
\\
\indent A\textit{ wavelet based} frequency and time variable (\textit{WFLVAR}) was constructed by taking ratios of parts of the decomposed continuous wavelet signal of the acoustic traces \cite{Alvine_thesis}. It replaced the older variable (\textit{FVAR}) used in the previous publication which took ratios of energy contained within select frequency regions of the Fourier transformed signals. 

\section{Acoustic Triangulation}
\label{Acoustic_Triangulation}
\begin{figure}[h]
\centering
\includegraphics[width=0.8\textwidth, origin=c, angle=0]{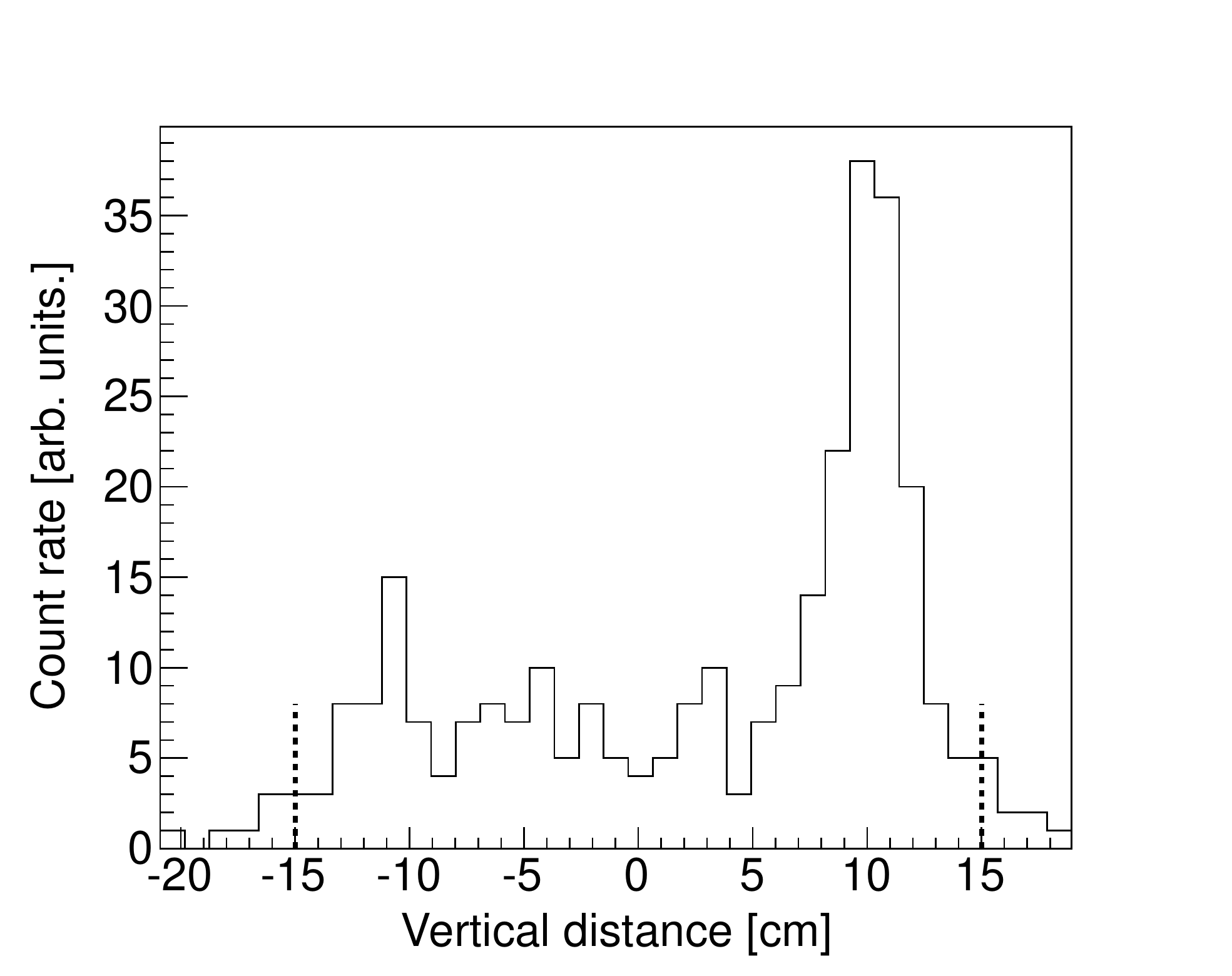}
\caption{Event localization by acoustic triangulation provides a powerful new tool for background discrimination. Shown here is the vertical profile of the count rate in one of the detector modules (\# 145). The center of the droplet emulsion is located at 0 cm, top and bottom at~$\pm$ 15 cm. A notable increase in $\alpha$-decay events shows up at the top, close to the interface between the droplet emulsion and the mineral oil buffer.}
\label{fig2}
\end{figure}

\indent Event localization by acoustic triangulation  and fiducialization are a new feature introduced in this analysis and   turned out to provide a powerful background discrimination especially against local alpha contamination and non-particle events happening at the container walls. This new technique uses the time differences between sound signals by different piezoelectric sensors to reconstruct the position of a bubble nucleation. First an event time for each piezoelectric sensor was found on raw and 18~kHz high-pass filtered signals from each piezoelectric sensor. Two methods were developed to find the event time: a comparison between two moving signal averages, averaged over 10~\textmu s and 45~\textmu s, respectively, and a cumulative shape indicator, where the cumulative time weighted average amplitude was compared with a uniform time integral to find the greatest separation, and hence the most likely signal start time. These arrival times were fit using a multi-parameter fit to obtain the localization point separately, giving four measurements of the bubble position. The results were then combined using a weighting that is inversely proportional to the fit quality $\chi^{2}$ \cite{Aubin_master}.
\\
\indent The performance was tested with 3 cm sized emulsion samples, suspended in a water filled detector module and a spatial resolution was found varying from $\pm$~0.8~cm in the center of the detector up to $\pm$ 2 cm at the walls. The localization uncertainty increases near the detector walls due to sound propagation effects; no events are lost, but some are reconstructed slightly outside of the physical detector volume.   The same measurements were used for a direct determination of the speed of sound in the emulsion. These measurements were complemented by calibration data in 12 detectors at temperatures from 30~$^{0 }$C to 40~$^{0 }$C and where the speed was added as an additional fit parameter. The measured mean value   v$_{s}$  = 1507~$\pm$~141~m/s at 40~$^{0 }$C agrees within uncertainties with the speed of sound in water v$_{s}$~=~1528.88~m/s which constitutes  78\% of the total mass of the detector.  
\\
\indent Applying acoustic localization in data taking runs, an overabundance of events was observed in seven modules at the top of the detectors, close to the interface between emulsion and the hydraulic fluid for compression (mineral oil). This increase was not present in calibration runs and remained constant with operating temperature, so it was inferred that these were ``hotspots" of alpha background contaminations on the surface, rather than an in-homogeneity in the droplet distribution. 
\\
\indent In order to cope with this type of background a fiducial volume has been defined by an iterative process. Starting with a central volume of 5 cm  radius and $\pm$ 8 cm in height, the count rate (in cts/gh) within this core volume was taken as a reference. Next, the active volume was gradually increased, as long as the count rate remained within one sigma of the core value, and this for all temperatures. An example is given in Fig. \ref{fig2} which shows the vertical profile of the count rate in one of the ``hot-spike" detector modules (\# 145). For these modules a tighter fiducial cut was implemented, reducing the background rate substantially. 
\\
\indent The event localization and wavelet analyses were also particularly useful for the discrimination of so called ``mystery events": for high temperature runs above ~45~$^{0}$C (i.e. low recoil threshold) a new type of background was observed in the data especially for seven detectors with decreased intrinsic alpha-background ($<$~10~cts kg$^{-1}$d$^{-1}$). This background was characterized by an increasingly large rate at high temperatures, similar in profile to a WIMP signal or neutron background. However, this background was not found in all detectors and when it was present had inconsistent and varying rates between modules. Data at 50~$^{0}$C particularly exhibited this class of background  events. By localizing the events it was noticed that they were concentrated at the edges of the detectors (both along the walls and at the top and bottom of the acrylic container). The most probable cause of these events are shear and stress effects at and in the vicinity of the emulsion interfaces. A typical fiducial cut of \textit{r $<$ }6 cm around the center of the detector, together with the wavelet analysis, was able to remove the mystery events altogether. This allowed the inclusion of the 1 keV threshold data for the first time. The active mass contained within the restricted fiducial volume was measured using the AmBe calibration source runs, and it was found that each detector had its fiducial mass reduced by this radial cut by about 30$\%$.
\\
\indent A summary of the final $^{19}$F fiducial masses and exposures used in  this analysis for each detector are given in Table \ref{Table_mass_exposure}. The integrated fiducial mass of the 32 detectors amounts to  1.41~$\pm$~0.11 kg of $^{19}$F and corresponds to 59.5~\% of the total fluorine mass.  
\\

\begin{table}[h]
\begin {center}
\begin{tabular}{l D{,}{\,\pm\,}{4} D{,}{\,\pm\,}{6} l D{,}{\,\pm\,}{4} D{,}{\,\pm\,}{3}}
\toprule
\multicolumn{1}{l}{Det}
& \multicolumn{1}{l}{Fid. Mass}
& \multicolumn{1}{l}{Exposure}
& \multicolumn{1}{l}{Det}
& \multicolumn{1}{l}{Fid. Mass}
& \multicolumn{1}{l}{Exposure}\\
\multicolumn{1}{l}{}
& \multicolumn{1}{l}{g(F)}
& \multicolumn{1}{l}{kg(F)d}
& \multicolumn{1}{l}{}
& \multicolumn{1}{l}{g(F)}
& \multicolumn{1}{l}{kg(F)d}
\\
\midrule
106 & 21.63,1.08 & 3.23,0.16 & 153 & 34.64,2.53 & 9.00,0.66\\
108 & 26.30,1.31 & 3.94,0.20 & 154 & 32.48,2.38 & 8.29,0.61\\
110 & 70.86,3.54 & 10.65,0.53 & 155 & 31.65,1.67 & 8.32,0.44\\
112 & 11.74,0.59 & 1.76,0.09 & 156 & 52.61,2.63 & 13.50,0.68\\
123 & 32.06,1.60 & 4.80,0.24 & 157 & 60.78,3.04 & 9.02,0.45\\
131 & 33.14,1.89 & 7.74,0.44 & 158 & 41.86,2.09 & 6.32,0.32\\
136 & 81.88,4.09 & 12.26,0.61 & 159 & 48.21,2.41 & 5.19,0.26\\
137 & 16.10,4.39 & 3.72,1.02 & 160 & 73.37,3.67 & 7.91,0.40\\
141 & 31.64,1.83 & 5.86,0.34 & 161 & 66.44,3.32 & 9.89,0.49\\
144 & 59.56,2.98 & 13.54,0.68 & 162 & 31.71,1.82 & 4.40,0.25\\
145 & 41.15,2.10 & 7.63,0.39 & 163 & 28.26,2.37 & 3.01,0.25\\
146 & 31.67,1.58 & 4.72,0.24 & 164 & 27.86,1.96 & 3.05,0.21\\
147 & 41.23,2.21 & 7.72,0.41 & 165 & 56.39,4.11 & 8.08,0.59\\
148 & 91.74,4.67 & 16.64,0.85 & 166 & 60.61,3.03 & 6.60,0.33\\
150 & 25.52,1.92 & 3.74,0.28 & 167 & 65.00,3.25 & 6.98,0.35\\
151 & 36.59,6.81 & 7.85,1.46 & 168 & 43.61,2.75 & 4.75,0.30\\
\bottomrule
\end{tabular}
\caption{Summary of the performance parameters of all detectors used in this analysis. The active masses refer to the mass content  of  $^{19}$F in a module after application of individual fiducial volume cuts.  Exposure values  cover data taken over the entire temperature range from 30~$^{0}$C $<$ T $<$ 50~$^{0}$C. The quoted mass errors are: 5$\%$ systematic uncertainty in the determination of the active mass and a 3$\%$ uncertainty by introducing the fiducial volume cut.}
\label{Table_mass_exposure}
\end {center}
\end{table}

\section{Analysis}
\label{Analysis}
\begin{figure}[h]
\centering
\includegraphics[width=0.8\textwidth, origin=c, angle=0]{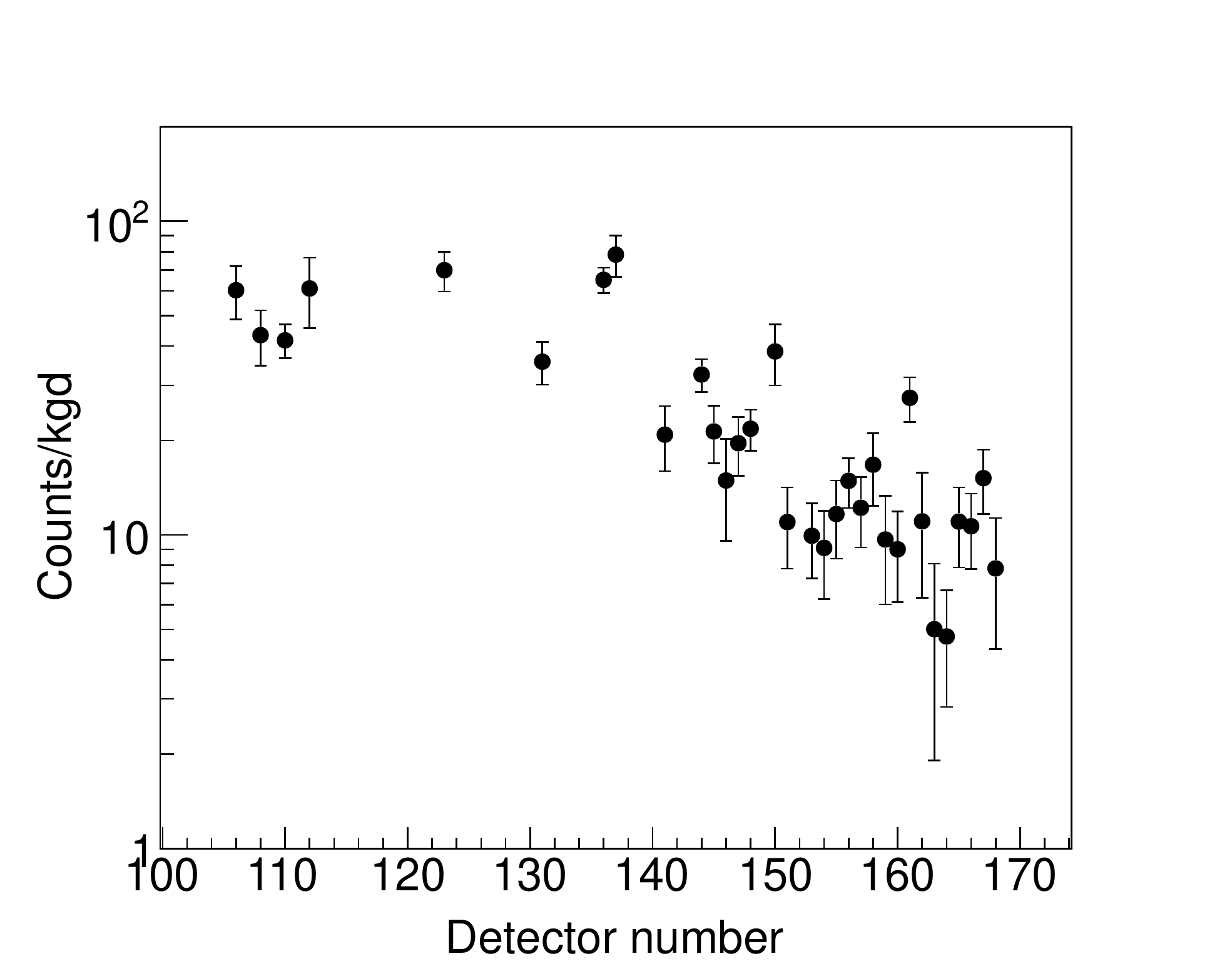}
\caption{Progressive reduction of background as a function of detector number. Shown is the count rate averaged over threshold energies in the range from 1 to 40 keV. The count rate after application of all cuts is flat in this region and indicative of the level of $\alpha$~-~contamination in each detector.}
\label{fig3}
\end{figure}

The selection of good runs and of true particle induced events above electronic and mechanical noise backgrounds proceeded in the following order: 
\\
\indent A list of golden runs was established for each detector. In order for a run to be good, at least six working acoustic readout channels were required; the duration of a run must have exceeded 15 hours and the gauge pressure in the detector had to be within 0.1 bar with respect to ambient pressure.
\\
\indent Two pre-selection cuts were applied to remove electronic noise artifacts from the data. Events were discarded when the pre-trigger noise region was found to be large and when the peak amplitude normalized to the pre-trigger noise region was found to be small. These cuts were found to only remove electronic noise and no efficiency correction was necessary. 
\\
\indent A time since last event cut was implemented to remove events thought to be caused by mechanical disturbances in the gel generated by fractures, deformation or gas bubble migration. The value used was 10 s during data taking runs and 0.1 s during calibration runs. The run exposure was corrected to account for this dead time.
\\
\indent After that the events had to pass the selections on \textit{EVAR}, \textit{RVAR}, \textit{QVAR}, \textit{TVAR} and \textit{WFLVAR}, with the cut values chosen such that a 95\% acceptance yield for calibration data was obtained. The event selection efficiency was estimated by accounting for variable correlations. The correlation matrix was measured from calibration runs and used as an  input to a pseudo Monte Carlo simulation. For each detector and temperature the efficiency was extracted by testing the number of simulated events that passed all cuts and a polynomial fit to the  efficiency was made. The fit value was used as the efficiency correction and was typically (detector and temperature dependent) in the range of 80 - 90 $\%$.
\\
\indent Finally the fiducial volume cut was applied as described in Section \ref{Acoustic_Triangulation}. The active mass was corrected to account for the reduction in exposure.
\\
\indent The effects of the applied cuts for two temperatures on the trigger rates are illustrated for detector module 153 in Table \ref{Table_Rate_per_cut}. For this module the fiducial volume cut plays an important role in removing non-particle induced events and and alpha particle ``hotspots".  
\\
\indent After correcting for cut acceptances and dead time the event rates for each detector at each temperature were normalized with respect to the active mass ($^{19}$F) and data taking  time. The count rates of all detectors showed a flat plateau in the range from 1.05 to 40 keV (50 - 28~$^{0}$C), similar to that observed in the presence of $\alpha$ - emitters (Fig. \ref{fig1}) in the droplets. The count rates averaged over the plateau range are shown in Fig. \ref{fig3} and are indicative of the level of $\alpha$ - contamination in the individual detectors, ranging from 80~cts~kg$^{-1}$d$^{-1 }$ for earlier modules to 5~cts~kg$^{-1}$d$^{-1}$ for the best of the last set of detectors.

\begin{table}[h]
\begin {center}
\begin{tabular}{l D{,}{\,\pm\,}{3} D{,}{\,\pm\,}{3} }
\toprule
Detector 153
& \multicolumn{1}{l}{30 $^{0}$C}
& \multicolumn{1}{l}{50 $^{0}$C}
\\
\midrule
Triggers/kgd & 241.5,8.4 & 5385.9,32.1\\
After burst cut & 26.9,3.8 & 700.4,49.2\\
After EVAR & 19.2,2.6& 32.62,3.4\\
After RVAR,  QVAR, TVAR & 19.1,2.6 & 31.2,2.7\\
After WFLVAR & 18.3,2.5 & 30.1,2.7\\
After fid. cut & 9.7,3.1 & 8.6,2.4\\
\bottomrule
\end{tabular}
\caption{Effect of the applied cuts on the count rate in detector 153 at 30 $^{0}$C and 50 $^{0}$C.}
\label{Table_Rate_per_cut}
\end {center}
\end{table}

\begin{figure}[h]
\centering
\includegraphics[width=0.8\textwidth, origin=c, angle=0]{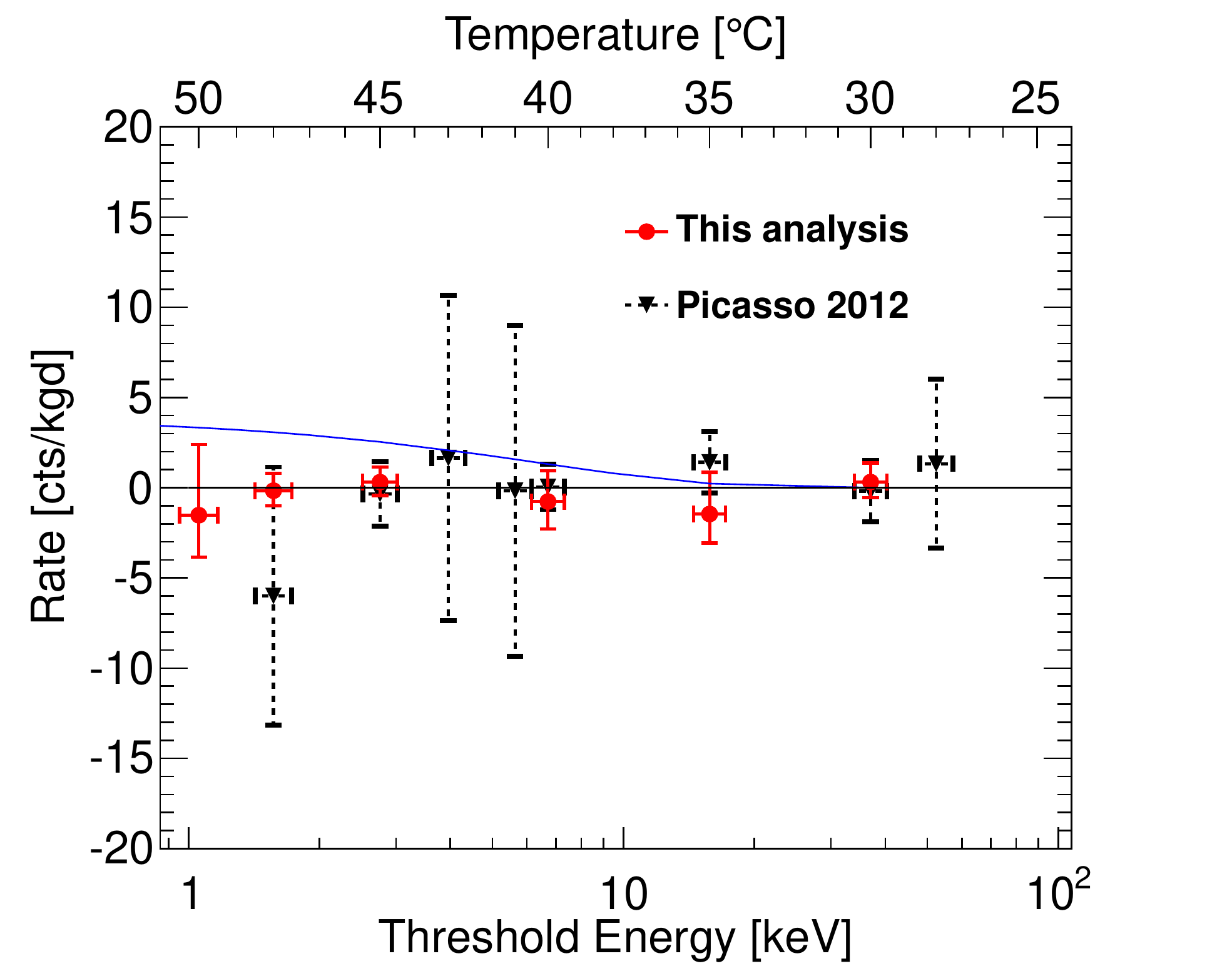}
\caption{Combined data from all detectors for WIMP runs for the present analysis. For each detector the average count rate was calculated over the temperature range 30~$^{0}$C~$<$~T~$<$~35~$^{0}$C and subtracted from individual data points at the higher temperatures.  Data for each detector and temperature are then combined in a weighted average. A hypothetical WIMP with M$_W$ = 15 GeV/c$^{2}$ and $\sigma^{SD}_p$ = 3.2  x 10$^{-2}$ pb is shown by the continuous curve (blue). PICASSO 2012 results are shown for comparison (black dotted).}
\label{fig4}
\end{figure}

\indent The progressive reduction in background was achieved by adding a 0.2~\textmu m  filtration stage for the monomer solution, by additional purification of the polymerizing agent (TEMED) and, for detectors 150 onward, by a doubling of all purification steps. In addition cover gas from LN$_{2}$  boil-off was used during all operations  to mitigate radon diffusion from ambient air into the emulsion ingredients.  This latter measure had no detectable effect on the detection sensitivity and threshold.   The origin of the residual $\alpha$ - background is  uncertain, but the acoustic signature of the events suggested that the activity was located primarily within the droplets. Detector 164 had the lowest background rate equivalent to a contamination level of 5$\times$10$^{-12}$ gU g$^{-1}$ in the C$_{4}$F$_{10}$ droplets. 
\\
\indent In order to combine the data of all detectors for illustrative purposes in a single plot, we adopted the following procedure: for each detector the average count rate was calculated over the temperature range from 30~$^{0}$C to 35~$^{0}$C where WIMPS with masses M$_{W}$ $<$ 15 GeV/c$^{2}$ do marginally contribute; this count rate was taken as an approximation of the $\alpha$ - background level of the detector and was subtracted from individual data points at different temperatures; the data for each detector and temperature were then combined in a weighted average; and finally, the temperatures were converted into threshold energies by taking into account that due to the somewhat elevated mine pressure (1.2 bar) the measured temperature at the location of the experiment corresponded to a threshold with a temperature at surface (where the calibration was performed), reduced by 2~$^{0}$C. The threshold dependence on pressure for a given temperature was measured by PICASSO with mono-energetic neutron test beams and is reported in \cite{PICASSO_NIM}.
\\
\indent 
The resulting threshold energy spectrum is shown in Fig. \ref{fig4}, where the error bars are dominated by statistics and reflect the time spent at each respective temperature. It is interesting to note that the count rates of all detectors as a function of recoil energy are essentially constant and that for modest changes in temperature from 30~$^{0}$C to 50~$^{0}$C the dynamic range in threshold energy sensitivity is large and covers the region from 1 keV up to 40 keV. No signal above background was observed.
\section{Results}
\label{re}

\begin{figure}[h]
\centering
\includegraphics[width=0.8\textwidth, origin=c, angle=0]{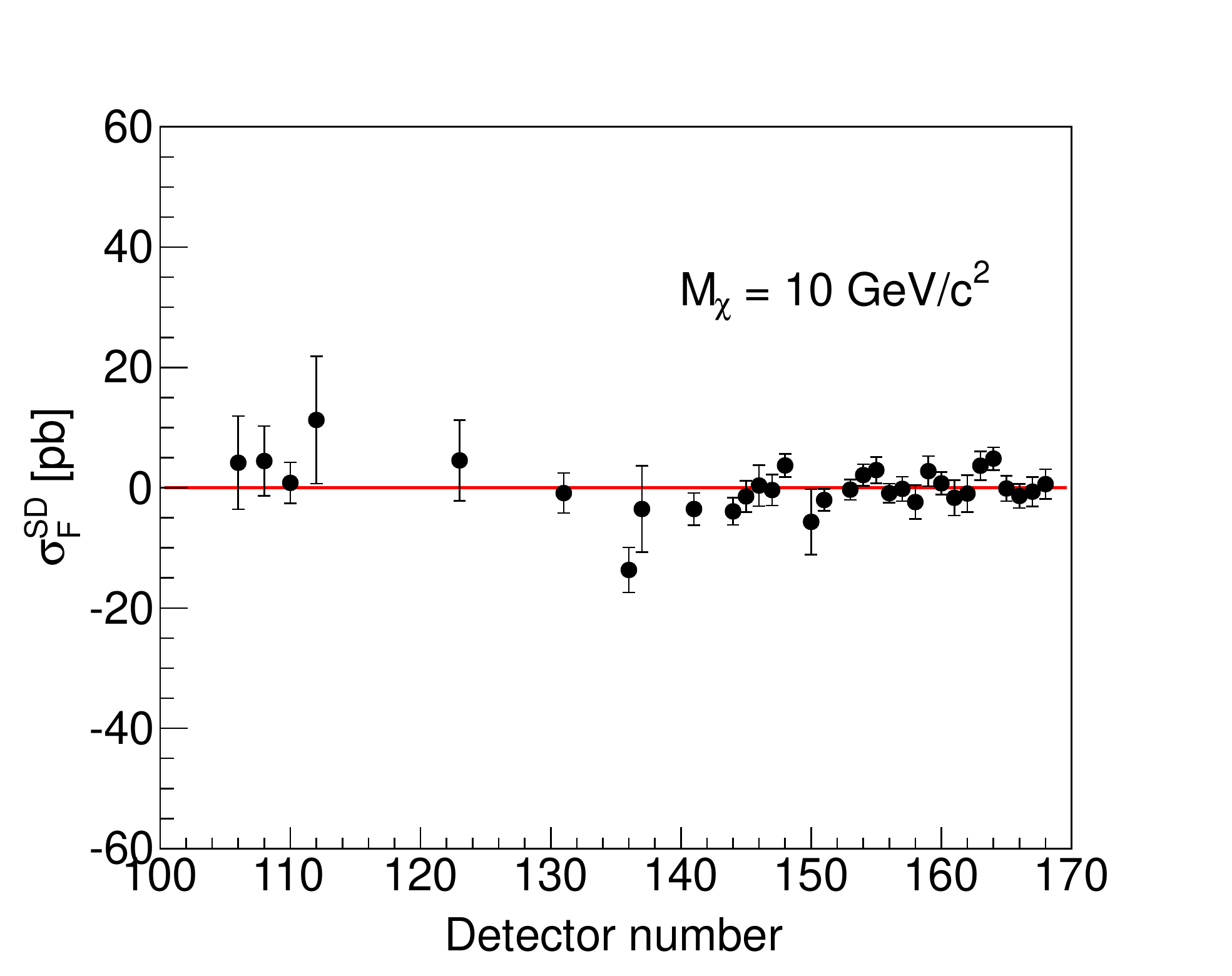}
\caption{Summary of the performance of all 32 detectors used in this analysis. Cross section values in pb for WIMP interactions on $^{19}$F are quoted for a resolution parameter \textit{a} = 5 and for a WIMP mass of 10 GeV/c$^2$, which is close to the maximum sensitivity. Systematic uncertainties are included as listed in the text. The detector number follows the time of fabrication. }
\label{fig5}
\end{figure}

 To search quantitatively for a dark matter signal the measured rates as a function of threshold energy have to be compared to those predicted for interactions of WIMPs in our galactic halo on $^{19}$F nuclei in the presence of a constant alpha background in each of the detectors. We use the formalism described in \cite{WIMP_HALO_INFO} which approximates the recoil energy spectrum by an exponentially falling distribution and we use the standard halo parameterization with $\rho_{D}$~=~0.3~GeVc$^{-2}$~cm$^{-3}$, \textit{v}$_{0 }$ = 220 km s$^{-1}$ and \textit{v}$_{Earth}$ = 232 km s$^{-1}$. Still following \cite{WIMP_HALO_INFO} we assume a nuclear form factor of \textit{F}(q$^{2}$) = 1, justified by the light fluorine nucleus and the small momentum transfers involved. 
\\
\indent Since our detectors operate as  threshold devices, the observed rate at a given recoil energy threshold \textit{E}$_{Rth}$(\textit{T}) is given by
\begin{equation}\label{eq:1}
R_{obs}(M_{W},\sigma_{F},E_{R_{th}}(T)) = \int_{ E_{R_{th}}(T)}^{E_{R_{max}}} P(E_{R}, E_{R_{th}}(T)) \frac{dR}{dE_{R}} dE_{R},
\end{equation}
where \mbox{\textit{P}(\textit{E}$_{R}$,~$E_{R_{th}}$(\textit{T}))} describes the effect of a finite resolution at threshold and \textit{dR}/\textit{dE$_{R}$} is the WIMP induced recoil energy spectrum; the integral extends to \textit{E}$_{R_{max}}$ which is the maximum recoil energy a WIMP can transfer at its galactic escape velocity of \textit{v}$_{esc}$ = 544 km s$^{-1}$. The shape of the threshold curve is described by \cite{PICASSO_NIM}:
\begin{equation}\label{eq:2}
P(E_{R}, E_{R_{th}}(T)) = 1 - \exp\left(a(T) \left(1- \frac{E_{R}}{E_{R_{th}}(T)}\right)\right) . 
\end{equation}
\indent The parameter \textit{a}(\textit{T}) describes the steepness of the energy threshold. It is related to the intrinsic energy resolution of the detection process and reflects the statistical nature of the energy deposition and its conversion into heat. It depends only on temperature and pressure and has to be determined \mbox{experimentally}. The larger \textit{a }is, the steeper the threshold is. Our measurements with alpha emitters with well defined, mono-energetic recoil nuclei ($^{210}$Pb) indicate a sharp threshold that can be described with \textit{a} $>$ 10 at 146 keV (Fig.~\ref{fig1}). Test beam measurements with mono-energetic neutrons at lower energies from 5 to 100~keV fall in the range of 1 $<$ \textit{a} $<$ 10, where decreasing energies  favor smaller \textit{a}. Poly-energetic AcBe neutron responses can be fit best with \textit{a} $\approx$~5. As in \cite{PICASSO2012} we adopt a principal value of \textit{a} = 5 and let the parameter vary within the interval 1 $<$ \textit{a }$<$ 7.5 when estimating the uncertainty.
\\
\indent For WIMP masses smaller \textit{M}$_{W}$ $<$ 500 GeV/c$^{2}$ the response curves differ in shape from the flat alpha background of each detector. By fitting the WIMP response curve and a flat alpha background, an upper bound on the WIMP-fluorine interaction cross section $\sigma_{F}$ is obtained for each individual detector. For a given mass \textit{M}$_{W}$ the two parameters of the fit are the cross section $\sigma_{F}$ and a scale factor describing the constant $\alpha$-background. As an example, the result for each detector is shown in Fig. \ref{fig5} for a WIMP mass of \textit{M}$_{W}$ = 10 GeV/c$^{2}$, the mass region of highest sensitivity.

The detector number follows the time of fabrication and the increasing sensitivity reflects the gradual reduction in alpha background by improvements in purification of the detector ingredients shown in Fig.~\ref{fig3}. Combined in a weighted average, the maximum sensitivity occurs for WIMPs in the mass region around \textit{M}$_{W}~$=~10~GeV/c$^{2}$ and with a cross section of $\sigma_F$~=~0.083~$\pm$~0.448~$\pm$~0.039~pb~(1$\sigma$). The systematic error contribution was estimated as: an overall 5\% uncertainty in the acceptance of the selection variables; a 3 \% uncertainty in the recoil detection efficiency inferred from the response to  $\alpha$-particles; a 5\% uncertainty in the determination of the active mass; a 3 \% uncertainty by introducing the fiducial volume cut; a 1~\% uncertainty from energy scale shifts due to temperature uncertainties during neutron beam calibrations; the uncertainties due to atmospheric pressure changes were estimated \textit{$<$}1 \%, similar to the uncertainty of the hydrostatic pressure change along the vertical profile of the detectors.
\textbf{ }\begin{figure}[H]
\centering
\includegraphics[width=0.8\textwidth, origin=c, angle=0]{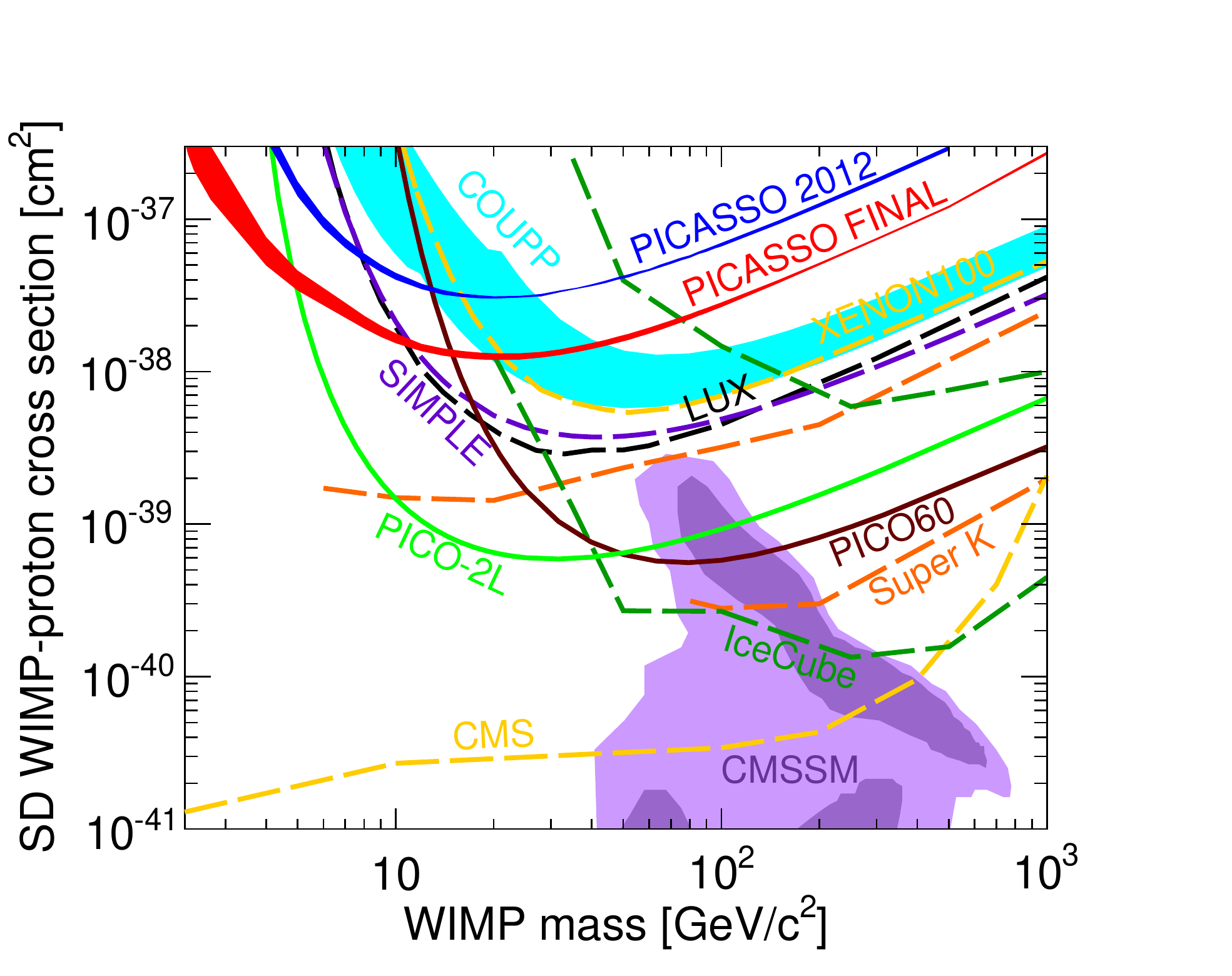}
\caption{Upper limits at 90$\%$ C.L. on SD-WIMP proton interactions. The final PICASSO limit is shown as a full red line along with limits from PICO-2L (green \cite{PICO2L_RUN2}), PICO60 (brown \cite{PICO60_CF3I}), COUPP-4 (light blue \cite{COUPP4_Erratum}), SIMPLE (dashed purple \cite{SIMPLE}), XENON100 (dashed light orange \cite{xenon2016}) and LUX (dashed black \cite{LUX2016_SD_Run3}). Indirect searches are represented by Ice-Cube (dashed dark green \cite{IceCube_2013}), SuperK (dashed orange \cite{SUPERK_2011,SUPERK_2015}) with comparable limits by ANTARES, Baikal and Baksan \cite{ANTARES,Baikal,Baksan}. Limits from accelerator searches by CMS are shown in dashed light orange \cite{CMS}. Comparable limits are set by ATLAS \cite{ATLAS}. The purple region represents predictions in the framework of the CMSSM \cite{CMSSM}.  } 
\label{fig6}
\end{figure}

Assuming that scattering of dark matter is dominated by spin-dependent interactions with the unpaired proton in $^{19}$F, the cross section $\sigma_{p}^{SD}$ for scattering on free protons is related to the measured cross section $\sigma_{F}$ by:
\begin{equation}\label{eq:3}
\sigma_{p}^{SD} = \sigma_{F} \left(\frac{\mu_{p}}{\mu_{F}}\right)^2 \frac{C_{p}^{SD}}{C_{p(F)}^{SD}}.
\end{equation}
Here $\mu_{p (F) }$ are the WIMP-proton (fluorine) reduced masses, \textit{C}$_{p}^{SD}$ is the spin enhancement factor for scattering on the free proton and ${C}_{p(F)}^{SD}$ is the corresponding quantity for scattering on protons in the $^{19}$F nucleus with the \mbox{ratio} ${C}_{p}^{SD}/{C}_{p(F)}^{SD}$ = 1.285 \cite{ref29,ref30}. The result for $\sigma_{F}$ is then converted with Eq.~\ref{eq:3} into a cross section on protons of $\sigma_{p}^{SD}$~=~(1.39~$\pm$~8.46~$\pm$~0.72)$\times$10$^{-3}$~pb (1$\sigma$;~\textit{a}~=~5), yielding a best limit of $\sigma_{p}$$^{SD}$~=~1.53$\times$10$^{-2}$ pb (90\% C.L.) for WIMP masses around 20 GeV/c$^{2}$. Adding the 114 kgd exposure of our 2012 data improves this limit slightly to $\sigma_{p}^{SD}$~=~1.32$\times$10$^{-2}$ pb (90\% C.L.) The resulting exclusion curve for the WIMP cross section on protons as a function of WIMP mass is shown in Fig.~\ref{fig6} together with published results in the spin-dependent sector. The broadening of the exclusion curve shows the effect of varying the energy resolution parameter \textit{a} within its uncertainty.
\textbf{ }\begin{figure}[H]
\centering
\includegraphics[width=0.8\textwidth, origin=c, angle=0]{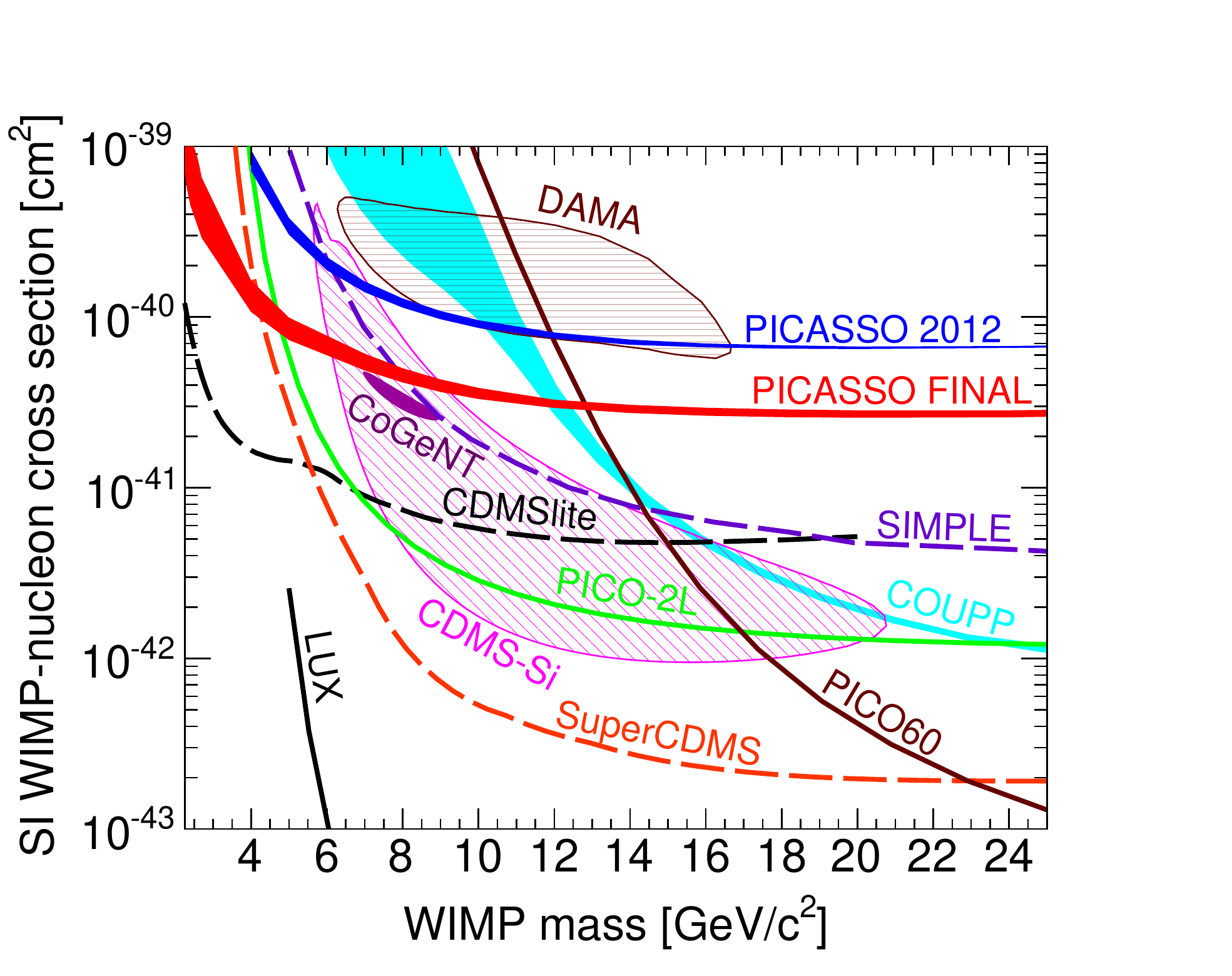}
\caption{Upper limits at 90$\%$ C.L. in the spin-independent sector. Only the region of interest in the range of low WIMP masses is shown. The closed contours are the allowed regions of DAMA (brown), CoGeNT (magenta) and CDMS-II Si (pink) \mbox{\cite{DAMA_LIBRA,CoGeNT,CDMS_SI_Signal}}. The final \mbox{PICASSO} limit is shown in full red, along with PICO-2L (green \cite{PICO2L_RUN2}), PICO60 (brown~\cite{PICO60_CF3I}), COUPP-4 (light blue \cite{COUPP4_Erratum}), SIMPLE (dashed violet~\cite{SIMPLE}), LUX (black~\cite{LUX2016SI_Run3Run4}), CDMSlite (dashed black \cite{CDMSlite}) and SuperCDMS (dashed orange \cite{SuperCDMS}). Similar limits (not~shown) are set by XENON10, XENON100, CRESST and PandaX-II \cite{XENON10,xenon2016,CRESST,PandaX}. }
\label{fig7}
\end{figure}
Similarly the limits on $\sigma_{F}$ can be translated into an upper bound on the WIMP nucleon cross section in the spin-independent sector with a maximum sensitivity at \textit{M}$_{W}$ = 20 GeV/c$^{2}$ and $\sigma_{p}^{SI}$ = 2.8$\times$10$^{-5}$ pb (90\% C.L.). A summary of allowed regions and exclusion limits is shown in Fig. \ref{fig7}. We are aware of recent efforts to treat WIMP-nucleon interactions in a more complete and model independent way by using the approach of effective field theories \cite{Effective_theory_DM} and it will be interesting to compare our $^{19}$F results with other targets within this broader analysis framework.

\section{Conclusions and Outlook}
\label{cno}

\indent PICASSO has operated a system of 32 superheated droplet detectors at SNOLAB with a combined exposure of 345.4 kgd. No indication of a WIMP signal was observed and a spin-dependent limit of~1.32$\times$10$^{-2}$ pb at \textit{M}$_{W}$~=~20~GeV/c$^{2}$ was set at a 90\% confidence limit. In the spin-independent sector  around 7 GeV/c$^{2}$, and close to the CoGeNT allowed region \cite{CoGeNT}, a limit of~4.9$\times$10$^{-5}$ pb (90\% C.L.) was extracted from the data. The use of the light target nucleus $^{19}$F, combined with an increased exposure at the low detection threshold of 1 keV resulted in increased leverage in the low WIMP mass region. The main improvements with respect to our previous published results are: a substantial reduction in intrinsic $\alpha$-background by up to a factor 10 in some modules and localization of events by acoustic triangulation.
\\
\indent The superheated droplet detector technique has proven to be a valuable technique for dark matter search especially in the spin-dependent sector and for low WIMP masses. The technique is easily scalable to multiple detectors; however the filling factor is only a few percent, the amount of surface area per active volume is much larger than in any other configuration of superheated liquid detectors and non-particle induced backgrounds start to become difficult to control in larger scale experiments. In addition the event by event $\alpha$-recoil discrimination using the acoustic signal energy discovered by PICASSO can be much easier applied in bulk superheated liquids. 
\\
\indent Without this important discrimination feature, a reduction in $\alpha$-activity by more than a factor of 10$^{3}$ would be required in order to obtain a sensitivity which equals that already reached by PICO-2L. This would not have been achievable with the existing purification technology for the components of the detector gel matrix. With this powerful acoustic discrimination tool the PICASSO group is now focused on large scale applications of superheated fluorinated halocarbons for dark matter detection using the more traditional bubble chamber technique as part of the broader PICO collaboration \cite{PICO2L_RUN2,PICO60_CF3I}.

\section*{Acknowledgements}
\label{ack}
We wish to express our thanks to SNOLAB and its staff for providing an 
excellent infrastructure, as well as technical support and advice 
whenever needed. We especially thank SNOLAB for providing the water 
tanks for the improved shielding of the experiment. Many thanks go to Tina Shepherd and Naomi Tankersley (IUSB) for fabricating the acoustic transducers. We wish to acknowledge the support of the National Sciences and Engineering 
Research Council of Canada (NSERC) and the Canada Foundation for Innovation 
(CFI). This work is also supported by the  National Science Foundation (NSF) under the Grants PHY-1205987 and PHY-1506377. We also acknowledge 
support from the Department of Atomic Energy (DAE), Govt. of India, 
under the project CAPP at SINP, Kolkata and the Czech Ministry of 
Education, Youth and Sports within the project MSM6840770029. We thank 
the members of the PICO collaboration for inspiring discussions and for 
providing the templates for the updated exclusion plots. 
\newpage
\section*{References}
\label{rf}
\bibliography{picasso_2016}
\bibliographystyle{model1-num-names}

\end{document}